\documentstyle[aps,epsfig,prl]{revtex}
\begin{document}
\title{Influence of a polaron dispersion and excitonic effect
on a magnetopolaron energy spectrum in a quantum well}
\author{ L. I. Korovin, I. G. Lang}
\address{A. F. Ioffe Physical-Technical Institute, Russian
Academy of Sciences, 194021 St. Petersburg, Russia}
\author { S. T. Pavlov\cite{byline1}}
\address {Escuela de Fisica de la UAZ, Apartado Postal C-580,
98060 Zacatecas, Zac., Mexico}
\twocolumn[\hsize\textwidth\columnwidth\hsize\csname
@twocolumnfalse\endcsname
\date{\today}
\maketitle \widetext
\begin{abstract}
\begin{center}
\parbox{6in} {
It has been shown, that a magnetopolaron discrete energy spectrum
is realized in the optical experiments where exciting light is
directed perpendicularly to a quantum well (QW) plane. The phonon
dispersion as well as Coulomb attraction can only shift the
magnetopolaron discrete energy levels and broad some of them.}
\end{center}
\end{abstract}

] \narrowtext
\section{Introduction.}
Let us consider  an energy spectrum of the electronic excitations
in a QW in a strong magnetic field directed perpendicularly to the
QW plane. So long as the system is homogeneous in the QW plane
($xy$ plane) any excitation (an exciton or electron-hole pair) can
be characterized by a quasi-momentum $\hbar{\bf {\cal K}}_\perp$
in the QW plane, if the wave functions are chosen properly.
 Such functions have been obtained earlier for the
excitons \cite{1} and magnetopolaron-hole pairs\cite{2}. The
quasi-momentum $\hbar{\bf {\cal K}}_\perp$ may be a quantum number
because the excitation, consisting of an electron and hole, is
neutral one. (Let us remind that in a strong magnetic field there
are not the states for an electron (hole) where  ${\bf {\cal
K}}_\perp$ would be a quantum number). Our aim is to determine in
principle how the magnetopolaron theory changes when a LO phonon
dispersion and excitonic effect are taken into account, i. e. a
Coulomb interaction between an electron, defining the polaron, and
hole, weakly interacting with LO phonons.

If the excitations are created in a QW by light, the condition
 $\bf {\cal K}_\perp=\bf {\kappa}_\perp$ must be satisfied, where
$\bf {\kappa}_\perp$ is the inplane projection of the light wave
 vector. One obtains ${\bf \cal K}_\perp=0$ if ${\bf
 \kappa}_\perp=0$. It is obviously that only discrete energy
 spectrum is possible at ${\bf \cal K}_\perp=0$ and in the case of
 a finite motion. In particular that means that at a normal light
 incidence neither the LO phonon dispersion, nor Coulomb interaction
 between an electron and hole results in a broadening of the energy
 levels into the bands. The mentioned above factors may  only
  shift  the discrete energy levels and change  the
 corresponding inverse lifetimes $\gamma$. Our results confirm
 this general statement. We have shown that the excitonic effect
 leads to the dependence of the energy of the magnetopolaron -hole pair
from the quasi-momentum $\hbar{\bf {\cal K}}_\perp$ of the
aggregate motion, what can be detected in experiments including an
oblique light incidence on the QW plane.

The magnetopolaron effect (the Johnson-Larsen effect) has been
discovered in \onlinecite{3,4,5} (see also the reviews
\onlinecite{6,7,8}). The magnetopolarons are created in 3D-systems
as well as in 2D-ones, for instance, in QWs. The distance between
the magnetopolaron energy levels in 3D-systems is $\sim
\alpha^{2/3}\hbar\omega_{LO}$\cite{9}, where  $\alpha$ is the
Fr\"ohlich non-dimensional electron-phonon coupling
constant\cite{10}, and in 2D-systems $\sim
\alpha^{1/2}\hbar\omega_{LO}$
\cite{11,12,13,14,15,16,17,18,19,20,21,22,23,24}.

\section{Influence of the phonon dispersion on the magnetopolaron
energy spectrum.}

The theory of magnetopolarons in a QW without taking into account
a phonon dispersion (i. e. in an approximation, where all the
phonons, taking part in a polaron formation, have the same
frequency $\omega_{LO}$) have been proposed in \onlinecite{19}.
The electrons interact with the confined and interface phonons. In
the continuum approximation \cite{25} (i. e. in the limit $a\to0$,
where $a$ is the lattice constant) the confined phonons have the
frequency $\omega_{LO}$ without a dispersion, but the interface
phonons have a dispersion, i. e. the dependence of the frequency
from the modulus $q_{\perp}$ of the inplane phonon wave vector.
The frequency of the interface phonons depends on the parameter
$q_{\perp}d$, where  $d$ is the QW's width. One has to take into
account the dependence of the phonon frequency from $q_{\perp}a$.
It has been shown in \onlinecite{23}, that in the case of the wide
QWs the approximation is applicable, in which the interaction with
the bulk  phonons is substituted for the interaction with the
confined phonons, and  the interaction with the interface phonons
may be neglected. Obviously, that in such a case we have only a
dispersion due to the deviation from the continuum model. In
\cite{14} the polarons are considered with taking into account the
interaction with the interface phonons.
 In \onlinecite{2,23} a classification of the magnetopolarons has
 been demonstrated. For example let us consider here the
 magnetopolaron $A$. It appears as a result of the crossover of
 the energy levels of the electron-phonon system with the indexes
$m, n=0, N=1$ and $m, n=1, N=0,$ respectively. $m$ is the
size-quantization quantum number, $n$ is the Landau quantum
number, $N$ is the number of LO phonons. The energies of the first
and second levels are
$\varepsilon_{m}^e+\Sigma_0+\hbar\omega_{LO}$ and $
\varepsilon_{m}^e+\Sigma_1$, respectively, where
$\varepsilon_{m}^e$ is the energy of the $m$-th size-quantized
energy level, measured from the QW's bottom. For instance, for the
QW with the infinitely high barriers
\begin{equation}
\varepsilon_{m}^e=\frac{\hbar^2{\pi^2}m^2}{2m_{e}d^2},
\end{equation}
where $m_e$ is the electron effective mass, $m=1,2,3\ldots .$ The
designations are introduced
\begin{equation}
\Sigma_0=\hbar\omega_{eH}/2,\quad
\Sigma_1=3\hbar\omega_{eH}/2,\quad\omega_{eH}=|e|H/(m_{e}c),
\end{equation}
e is the electron charge, H is the magnetic field intensity, $c$
is the light velocity  in a vacuum. We do not take the phonon
dispersion into account up to now. Obviously the energy levels
cross over when
\begin{equation}
\omega_{eH}=\omega_{LO}.
\end{equation}
When the resonance condition Eq. (3) is satisfied, the role of the
electron-phonon interaction increases sharply, what leads to the
splitting of the energy levels of the electron-phonon system and
to the magnetopolaron formation.

The theory, which has been developed in \onlinecite{19}, is
applicable when the QW's widths are not too wide and
,consequently, the distance between the magnetopolaron levels is
small in comparison to the distance between the size-quantized
energy levels. It has been shown in \onlinecite{23} that the last
condition is satisfied for a QW of the system AlSb/GaAs/AlSb at
$d\leq 500A$. The low temperatures are supposed when the optical
phonons are not excited.

We consider a rectangular QW of the I type with an energy gap
$E_g$. The magnetic field is directed along the axis $z$
perpendicularly to the QW's plane, the vector-potential is  ${\bf
A}={\bf A}(0,xH,0)$. The electron wave function in the QW is
\begin{equation}
\Psi_{n,k_{y},m}^{e}(x,y,z)=
\Phi_{n}(x+a_{H}^{2}k_{y})\frac{1}{\sqrt
L_y}e^{ik_{y}y}\varphi_m^e(z),
\end{equation}
where
$$\Phi_n(x)=\frac{e^{-x^2/2a_H^2}H_n(x/a_H)}{\sqrt{\pi^{1/2}2^nn!a_H}},
~~ a_H=\sqrt{\frac{c\hbar}{|e|H}},$$ $H_n(t)$ is the Hermitian
polynomial, $L_y$ is the normalization length, $\varphi_m^{e}(z)$
is the real electron wave function, corresponding to the $m$-th
size-quantized energy level (see, for instance, \onlinecite{19}).
The electron-phonon interaction is written as
\begin{equation}
V=\sum_\nu[{\cal C}_\nu({\bf r}_\perp,z)b_\nu+{\cal
C}_\nu^\ast({\bf r}_\perp, z)b_\nu^+],
\end{equation}
where $\nu$ is the set on indexes, consisting of
 $\bf {q}_\perp$
and other indexes $j$, which characterize the confined and
interface phonons; $b_\nu^{+}~(b_\nu)$ is the phonon creation
(annihilation) operator,
\begin{equation}
{\cal C}_\nu({\bf r}_{\perp},z)= C_{\nu}e^{i{\bf q}_ {\perp}{\bf
r}_\perp}\eta_{\nu}(z);
\end{equation}
the values $C_{\nu}\eta_\nu(z)$ for the electron-phonon
interaction with phonons are determined in \onlinecite{25}.

As it has been shown in \onlinecite{23}, at $d\geq200A$ in GaAs
the application of the electron-bulk phonon interaction is a good
approximation. In this approximation the index set $\nu$ transits
into $\bf q$, where ${\bf q}=({\bf q}_{\perp},q_z)$ is the 3D
phonon wave vector and, according to \onlinecite{10},
\begin{equation}
\eta_\nu(z)=e^{iq_zz}, C_{\nu}=C_{q}=-i\hbar\omega_{LO}
\left(\frac{4\pi{\alpha}l^3}{V_0}\right)^{1/2}\frac{1}{ql},
\end{equation}
where $$l=\sqrt\frac{\hbar}{2m_e{\omega_{LO}}},\quad
\alpha=\frac{e^2}{2\hbar\omega_{LO}l}(\frac{1}{\varepsilon_\infty}-
\frac{1}{\varepsilon_0}),$$ $V_0$ is the normalization volume,
$\varepsilon_0(\varepsilon_\infty)$ is the static (high frequency)
dielectric function \cite{10}. For GaAs $\alpha\simeq0.071,$
$l\simeq40A$. Applying Eqs. (4) and (6),one finds that the
interaction matrix elements are equal
\begin{eqnarray}
\int d^{3}r\Psi_{n^\prime,k_y^\prime,m}^{e\ast}{\cal C}_
\nu^\ast({\bf r}_\perp ,z)\Psi_{n,k_y,m}^{e}=
U_{n,n^\prime}^\ast(\nu)\nonumber\\ \times e^{(ia_H^2q_x(k_y+k_y^
\prime)/2)}\delta_{k_y^\prime,k_y-q_y},
\end{eqnarray}
where the designations are introduced:
\begin{equation}
U_{n,n^\prime}^\ast(\nu)=C_\nu^\ast{\cal K}_{n,n^\prime} (a_H{\bf
q}_\perp\times{\bf H}/H){\cal M}^\ast(\nu),
\end{equation}
\begin{eqnarray}
 K_{n,n^\prime}({\bf s})=\sqrt\frac{min(n!,n^\prime!)}{max(n!,n^\prime!)}
i^{|n-n^\prime|}\left(\frac{s}{\sqrt
2}\right)^{|n-n^{'}|}\times\nonumber\\ exp(-\frac{s^2}{4})
exp[i(\phi-\pi/2)(n-n^\prime)]L_{min(n,n^\prime)}^{|n-n^\prime|}(s^2/2)
\end{eqnarray}
\begin{equation}
{\cal
M}(\nu)=\int_{-\infty}^{\infty}dz[\varphi_m^e(z)]^2\eta_\nu(z),
\end{equation}
${\bf s}$ is the 2D vector, $s=\sqrt{s_x^2+s_y^2}$,
$\phi=arctg(s_y/s_x)$, $L_m^n(t)$ is the Laguerre polynomial.
Without taking into account the phonon dispersion \cite{1} the
following expression for the energy $E$ of the polaron $A$ has
been obtained:
\begin{equation}
E-\Sigma_1-\frac{\sum_\nu|U(\nu)|^2}{E-\Sigma_0-\hbar\omega_{LO}}=0.
\end{equation}
 Here $U(\nu)=U_{1,0}(\nu)$. Applying (10)
and(11), one obtains
\begin{equation}
|U(\nu)|^2=|C_\nu|^2~(a_H^2q_\perp^2/2)~exp(-a_H^2q_\perp^2/2)~|
{\cal M}(\nu)|^2.
\end{equation}
Resolving Eq. (12), one obtains
\begin{eqnarray}
E_p&=&\frac{1}{2}(\Sigma_1+\Sigma_0+\hbar\omega_{LO})\nonumber\\&\pm&
\sqrt{\frac{1}{4}(\Sigma_1-\Sigma_0-\hbar\omega_{LO})^2
+\sum_\nu|U(\nu)|^2},
\end{eqnarray}
where the index $p$ designates the magnetopolaron energy levels:
 $p=a$ corresponds to the upper level
(the sign minus in the RHS of Eq. (14)), $p=b$ corresponds to the
lower level (the sign minus in the RHS of Eq. (14))). The solution
Eq. (14) is right in the resonance vicinity Eq.
 (3). In the resonance sharply
\begin{equation}
E_p^{res}=\Sigma_1\pm\sqrt{\sum_\nu|U(\nu)|^2},
\end{equation}
thus, the polaron splitting equals
\begin{equation}
\Delta E^{res}=E_a^{res}-E_b^{res}=2\sqrt{\sum_\nu|U(\nu)|^2}.
\end{equation}
The magnetopolaron $A$ wave functions for $p=a, b$ have been
obtained in \onlinecite{23} .

Taking into account the phonon dispersion and applying the method
of \onlinecite{19}, we obtain
\begin{equation}
E-\Sigma_1-\sum_{{\bf q}_\perp,~j}\frac{|U({\bf q}_\perp,j)|^2}
{E-\Sigma_0-\hbar\omega_j({\bf q}_\perp)}=0
\end{equation}
instead of Eq. (12).

 Let us consider a quite wide QW, where an
approximation of interaction with the bulk phonons is applicable.
We determine the phonon dispersion as following (an anisotropy of
the phonon energy spectrum is neglected)
\begin{equation}
\omega_{LO}({\bf q})=\omega_{LO}-
\Delta\omega_{LO}(q),\qquad\Delta\omega_{LO}(q=0)=0.
\end{equation}
Then Eq. (17) takes the view
\begin{equation}
F(E){\equiv}E-\Sigma_1-\sum_{\bf q}\frac{|U({\bf q}_\perp,q_z)|^2}
{E-\Sigma_0-\hbar\omega_{LO}+\hbar\Delta\omega_{LO}(q)}=0,
\end{equation}
and, according to Eqs. (7), (11), (13),
\begin{eqnarray}
|U({\bf q}_\perp,q_z)|^2&=&\hbar\omega_{LO})^2 {4\pi\alpha l\over
V_0q^2}\nonumber\\&\times&
(a_H^2q_\perp^2/2)~exp(-a_H^2q_\perp^2/2)~|{\cal M}(q_z)|^2;
\end{eqnarray}
\begin{equation}
{\cal
M}(q_z)=\int_{-\infty}^{\infty}dz[\varphi_m^e(z)]^2exp(iq_zz).
\end{equation}
The function $F(E)$ has been calculated for the case of the
square-law dispersion
\begin{equation}
\Delta\omega_{LO}(q)=cq^2
\end{equation}
and under the resonance condition of Eq. (3).

The integral had been taken as a main value for those energies
$E$, when the denominator could equal 0. The function $F(E)$ is
represented in Fig. 1 in the dispersion absence for the cases:
$c=0$,( the curve 1), $c/gl^2=0.04$ (the curve 2), $c/gl^2=0.2$
(the curve 3), $g=\alpha^{1/2}\hbar\omega_{LO}.$ The curves 2 and
3 cross over the abscissa axis in the points
 $E_{b}^\prime$, $E_{c}$ and
$E_{a}^\prime$, obtained with taking into account the LO phonon
dispersion. The crossover points $$E_b=\Sigma_1-\sqrt
{\sum_\nu|U(\nu)|^2},\quad E_a=\Sigma_1+\sqrt
{\sum_\nu|U(\nu)|^2}$$ correspond to the theory without the LO
phonon dispersion. The differences  $E_{a}-E_{a}^\prime$ and
$E_{b}^\prime-E_b$ increase with growing of the dispersion
parameter as it is seen in Fig. 1. The small shifts of the polaron
levels correspond to the weak dispersion. The third crossover
point (to which the energy $E_c$ corresponds) appears only with
taking into account the phonon dispersion. A discussion of the
last result follows below.

The phonon dispersion may lead to the additional contributions
into the inverse lifetimes of the polaron states\cite{byline2}.
That can be explained with the help of the schematic Fig. 2, where
the energy levels  $E_{b}^\prime$, $E_{c}$ and $E_{a}^\prime$ are
represented together with the schematic curves, depicting the
dependence of the value
 $\Sigma_{0}+\hbar\omega_{LO}(q)$ on  of the 3D
 phonon wave vector under the resonant condition of Eq.
(3). Fig. 2b corresponds to the larger phonon dispersion, than
Fig. 2a. In Fig. 2b the curve $\Sigma_{0}+\hbar\omega_{LO}(q)$
does not cross over the energy levels
 $E_{a}^\prime$ and $E_{b}^\prime$.

That means, that the denominators
$E_{a}^\prime-\Sigma_{0}-\hbar\omega_{LO}(q)$ and
$E_{b}^\prime-\Sigma_{0}-\hbar\omega_{LO}(q)$ in the LHS of Eq.
(19) do not equal 0 and the real solutions $E_{a}^\prime$ and
$E_{b}^\prime$ are precise ones.

 Applying the method of \onlinecite{19} one can show, that the
 magnetopolaron wave functions, corresponding to the precise
 solutions
$E_{a}^{'}$ and $E_{b}^{'},$  in the resonance vicinity of Eq. (3)
have the view
\begin{eqnarray}
\Theta_{p,k_{y}}|0\rangle&=&\left[1+\sum_\nu\frac{|U(\nu)|^2}
{E_p^\prime-\Sigma_{0}-\hbar\omega_{LO}(\nu)}\right]^{-1/2}\nonumber\\
&\times&\left[\Psi_{1,k_{y},m}^{e}
+\sum_\nu\frac{exp[ia_{H}^{2}q_x(k_{y}-q_{y}/2)]}
{E_p^\prime-\Sigma_{0}-\hbar\omega_{LO}(\nu)}\right.\nonumber\\
&\times&\left. U^\ast(\nu)
\Psi_{0,k_{y}-q_{y},m}^{e}b_\nu^{+}\right]|0\rangle,
\end{eqnarray}
where $|0\rangle$ is the phonon vacuum wave function; $p$ equals
$a$ or $b$. The functions of Eq.(24) distinguish on the
corresponding functions without the phonon dispersion only by the
substitution $E_{p}$ by $E_{p}^\prime$ and $\omega_{LO}$ by
$\omega_{LO}(\nu)$ \cite{19}. The wave functions are orthogonal
and normalized, i. e.
\begin{equation}
\int
d^3r<0|{\Theta^{+}_{p^{\prime}k_{y}^\prime}}{\Theta_{pk_{y}}}|0>=
\delta_{p,p^\prime}\delta_{k_{y},k_{y}^\prime}.
\end{equation}
The orthogonalization of the wave functions Eq. (23) with indexes
$a$ and $b$ can be checked easily if one takes into account the
interrelation
\begin{equation}
\sum_\nu\frac{|U(\nu)|^2}{[E_a^\prime-\Sigma_0-\hbar\omega_{LO}(\nu)]
[E_b^\prime-\Sigma_0-\hbar\omega_{LO}(\nu)]}=-1,
\end{equation}
which can be obtained, if in the LHS of Eq. (19) one substitutes
$E_{a}^\prime$, afterwards $E_{b}^\prime$ and subtracts the second
expression from the first one.

 As far as the energy $E_{c}$ value is concerned, it is seen in Fig. 2
 , that the curve
$\Sigma_{0}+\hbar\omega_{LO}(q)$ crosses over always with the
energy level $E_{c}$, because this level is as closer to the
energy
 $\Sigma_1=\Sigma_{0}+\hbar\omega_{LO}$, as the phonon dispersion
 is weaker.
That means that the denominator
$E_{c}-\Sigma_{0}-\hbar\omega_{LO}(q)$ in the  LHS of Eq. (19)
equals 0 at some absolute value  $q_{c}$ of the phonon wave
vector. Consequently, the real value
 $E_{c}$ is not a precise solution of Eq.
(19).

In Fig. 2b the curve $\Sigma_{0}+\hbar\omega_{LO}(q)$ crosses over
not only the energy level
 $E_{c}$, but  the lower polaron level $E_{b}^\prime$ also.
It follows from this fact, that only  real solution $E_{a}^\prime$
is precise one, but the solutions, corresponding to the energy
levels $a$ and $b$, must contain some imaginary parts. That means
that the states $a$ and $b$ have the finite lifetimes, which we
designate as $\gamma_c^{-1}$ and $\gamma_b^{-1}$.

Let us try to calculate $\gamma_{c}$ and $\gamma_{b}$. We have to
generalize Eq. (19) so, that it could admit the complex solutions.
The generalization suppose the substitution of the desirable
energy $E$ by $E+i\delta$, where $\delta\to+0$. Then some
imaginary
 term will appear  in Eq. (19), connected with the circuition
of the integrand pole (in fact, the function $F(E+i\delta)$ is a
denominator of an one-particle retarded electron Green function).
Let us suppose, that the inverse lifetime
 $\gamma_{p}$ of the state $p$ is very small.
 Then,
adopting $\tilde{E}_{p}=E^\prime_{p}-i\hbar\gamma_{p}/2$, where
$E^\prime_{p}$ is the real value, and applying a decomposition
 on the small value $\gamma_{p}$,
 one obtains from Eq. (19) in a zero approximation
\begin{equation}
E_{p}^\prime-\Sigma_{1}-Re\sum_{\bf q}\frac{|U({\bf
q}_\perp,q_z)|^2}
{E_{p}^\prime-\Sigma_0-\hbar\omega_{LO}(q)+i\delta}=0,
\end{equation}
where $\delta\to +0$, and in the next approximation one obtains
\begin{eqnarray}
\gamma_p=-2Im\frac{1}{\hbar}\sum_{\bf q}\frac{|U({\bf
q}_\perp,q_z)|^2}
{E_{p}^\prime-\Sigma_0-\hbar\omega_{LO}(q)+i\delta}=\nonumber\\
=\frac{2\pi}{\hbar}\sum_{\bf q}|U({\bf q}_\perp,q_z)|^2
\delta[E_{p}^\prime-\Sigma_0-\hbar\omega_{LO}(q)].
\end{eqnarray}

Having the values $\gamma_p$, we can find, if they are small
indeed. Thus, we can check, if the method, descending to the Eqs.
 (26), (27), is applicable indeed. We will see below, that the
 method is applicable to the energy level
$b$ at the weak  phonon dispersion, but inapplicable to the energy
level $c$.

Let us substitute Eq. (20) into the RHS of Eq. (27) and use the
dispersion Eq. (22). In the calculations of the function
 ${\cal M}(q_z)$ we use the wave functions
\begin{equation}
\varphi_{m}^{e}(z)=(2/d)^{1/2}sin(m{\pi}z/d),~~~~~0<z<d
\end{equation}
(and $\varphi_{m}^{e}(z)=0$ outside this interval), corresponding
to the QW with the infinite barriers. One obtains
\begin{equation}
|{\cal M}(q_{z})|^2\equiv f_{m}(Q)=\frac{2(2\pi
m)^4(1-cosQ)}{Q^2[Q^2-(2\pi m)^2]^2},
\end{equation}
where $Q=q_zd$, $m$ is a number of the size-quantized energy
level. Integrating  the RHS of Eq. (27) on $q_\perp$ with the help
of the   $\delta$-function, let us represent $\gamma_p$ as an
integral on the variable $Q$
\begin{equation}
\gamma_{p}=\frac{2\alpha
l\hbar\omega_{LO}^2}{d(\Sigma_0+\hbar\omega_{LO}-E_{p}^{'})}
\int_0^{q_p d}dQe^{-x}xf_m(Q),
\end{equation}
where $E_{p}^{'}$ is the $p$ polaron level energy, calculated,
according to Eq. (26), with taking into account the phonon
dispersion,
\begin{equation}
q_{p}=\sqrt\frac{\Sigma_{0}+\hbar\omega_{LO}-E_{p}}{c},
\end{equation}
\begin{equation}
x=\frac{a_H^2q_p^2}{2}-\frac{Q^2}{\beta_0^2},
\end{equation}
\begin{equation}
\beta_{0}=\frac{\sqrt 2d}{a_H}.
\end{equation}

If the condition of Eq. (3) is satisfied, $\beta_{0}=d/l$. At any
value  $m$
\begin{equation}
\int_0^{\infty}dQf_m(Q)=3\pi/2.
\end{equation}
Therefore the integral in the RHS of Eq. (30) is always smaller
than $3\pi/2$. If the dispersion is very weak
\begin{equation}
dq_{p}>>1,~~~~~a_Hq_p>>1,
\end{equation}
the integral
\begin{equation}
\int_0^{q_p d}dQe^{-x}xf_m(Q)<<1.
\end{equation}
That means, that for the energy level $p=b$ the value $\gamma_p\to
0$ when the dispersion parameter $c\rightarrow 0$, because the
value $\Sigma_{0}+\hbar\omega_{LO}-E_{p}^\prime$, which is in the
denominator of the RHS of Eq. (30), tends to
$\Sigma_{0}+\hbar\omega_{LO}-E_{b}$. In Fig. 3 the position of the
energy level  $E_{a}^\prime$ is represented, as well as the
broadening of the energy level $E_{b}^\prime$ as a function of the
dispersion parameter $c$. One can see, that at the small values
$c$ the broadening is small and the approximate expression Eq.
 (30) for $\gamma_{b}$ is right. However, with the increasing
 $c$ the value $\gamma_{b}$
 increases so strongly, that the solutions of Eqs.
(26) and (27) become incorrect. For the sake of comparison let us
represent the expression for the magnetopolaron splitting $\Delta
E^{res}$, which has been obtained in \onlinecite{23} for the wide
QWs  in the limit
 $\beta_{0}>>2\pi m$:
\begin{equation}
\Delta E^{res}=\alpha^{1/2}\hbar\omega_{LO}\sqrt {6l/d}.
\end{equation}
Comparing Eq. (30) to Eq. (37), one finds, that
$\hbar\gamma_{b}<<\Delta E^{res}$ at a weak dispersion.

As far as the energy level $c$, according to Eq. (30), its
inversion lifetime $\gamma_c$ increases with the dispersion
decreasing. Indeed, in the denominator of the RHS of Eq. (30)
there is the value
 $\Sigma_{0}+\hbar\omega_{LO}-E_c$, which tends to 0 with
 decreasing the dispersion parameter and the value
$\gamma_c\rightarrow\infty$. That means that the method of the
sequential approximations of Eqs. (26)-(27) is inapplicable to
analyse the energy level $c$. The question about an existence of
this level is opened up to now.

\section{Influence of the excitonic effect on the magnetopolaron energy
spectrum.}

In the previous section we have examined a magnetopolaron, which
has been formed by an electron. In this section we consider some
magnetopolaron-hole pair. The influence of the Coulomb forces on
the energy spectrum of an electron-hole pair (EHP) is weak under
conditions
\begin{equation}
a_{exc}^{2}>>a_H^2,~~~~~~~~~~~~~~a_{exc}>>d,
\end{equation}
where $a_{exc}=\hbar^2\varepsilon_{0}/\mu e^2$ is the radius of
the Wannier-Mott exciton in a magnetic field absence,
$\mu=m_{e}m_{h}/(m_{e}+m_{h})$ is the reduced effective mass.
Applying the parameters $m_{e}/m_{0}=0.065, m_{h}/m_{0}=0.16,
\varepsilon_{0}=12.55, \hbar\omega_{LO}=0.0367$ one obtains for
GaAs
\begin{equation}
a_{exc} = 146A,~~~~~~~~~~     a_{H}^{res} = 57.2A,
\end{equation}
$a_{H}^{res}=\sqrt{c\hbar/(|e|H_{res})}$, $H_{res}$ is the
magnetic field, corresponding to the magnetopolaron resonance Eq.
 (3),
$m_0$ is the bare electron mass. $H_{res} = 20.2T$ for  GaAs. One
obtains from Eq. (39), that
 $(a_{H}^{res}/a_{exc})^2\simeq 0.154$,
i. e. the first of the conditions of Eq. (38) is satisfied,
however the second condition demands to consider the QWs with the
widths $d<<146A$, i. e. more narrower than we have considered in
the previous section.

The first inequality of Eq. (38) is equivalent to the following
one
\begin{equation}
\hbar\omega_{\mu H}/2 >> \Delta E_{exc},
\end{equation}
where $\omega_{\mu H}=|e|H/\mu c$ is the cyclotron frequency,
$\Delta E_{exc} = \hbar^{2}/\mu a_{exc}^2$ is the exciton coupling
energy in a magnetic field absence.

Under condition Eq. (38) the Coulomb interaction of an electron
and hole may be considered as a weak perturbation and one can
calculate the first order corrections to the EHP energy according
to the perturbation theory  (see \onlinecite{1}, where the 2D case
has been considered).

The EHP unperturbed wave functions are chosen as the wave
functions \cite{2} with the index  ${\bf {\cal K}}_{\perp}$ and
indexes $n$ and $n^\prime$, corresponding to the relative motion
of the electrons and holes \cite{26}. Let us note, that the
indexes
 n and $n^\prime$
are connected single-valuedly with the Landau quantum numbers
$n_e$ and $n_h$ of electrons and holes, respectively: at
$n_{e}>n_{h}~~n=n_{h}, n^\prime=n_{h}-n_{e}<0$, but if
$n_{e}<n_{h}$, then $n=n_{e}, n^\prime=n_{h}-n_{e}>0$. The EHPs
with taking into account the Coulomb forces, which can be called
as excitons, are characterized by the same sets of indexes ${\bf
{\cal K}}_{\perp},n,n^\prime$ and
 ${\bf {\cal K}}_{\perp},n_{e},n_{h}$.
That has been shown in \onlinecite{1} that the corrections to the
EHP energy due to the Coulomb interaction, depend on the inplane
quasi-momentum  $\hbar{\cal K}_\perp$ in the QW plane, i. e. the
Coulomb forces lead to the exciton dispersion.

The energy corrections due to the excitonic effect may be
represented as two parts. The first depends only on the indexes
$n_e$ and $n_h$ and corresponds to ${\bf {\cal K}}_\perp=0$.

The second (the rest part) depends on ${\bf {\cal
K}}_{\perp},n_{e},n_h$ and describes the exciton dispersion. The
exciton energy, characterized by the quasi-wave vector
 ${\bf{\cal K}}_\perp$, and consisting of the electron with the indexes
$n_e$ and $m_e$ and the hole with the indexes $n_h$ and $m_h$,
where
 $m_{e}(m_{h})$ is the number of the size-quantized energy level,
 is equal
\begin{eqnarray}
{\cal E}_{n_{e},{n_h},m_{e},m_{h}}({\cal
K}_\perp)=E_{g}+\varepsilon_{m_{e}}^{e}+
\varepsilon_{m_{h}}^{h}\nonumber\\+(n_{e}+1/2)\hbar\omega_{eH}\nonumber\\
+(n_{h}+1/2)\hbar\omega_{hH}+ \Delta {\cal E}_{n_{e},n_{h}}({\bf
{\cal K}}_\perp),
\end{eqnarray}
where $\Delta{\cal E}_{n_{e},n_{h}}({\bf {\cal K}}_\perp)$ is the
Coulomb correction to the exciton energy. Let us separate the
contribution at ${\bf {\cal K}}_\perp=0$:
\begin{equation}
\Delta {\cal E}_{n_{e},n_{h}}({\bf {\cal K}}_\perp)= \Delta {\cal
E}_{n_{e},n_{h}}({\bf {\cal K}}_{\perp}=0)+ \Delta_{1} {\cal
E}_{n_{e},n_{h}}({\bf {\cal K}}_\perp).
\end{equation}
The exciton states with indexes $n_{e},n_{h},m_{e},m_{h},{\bf
{\cal K}}_\perp$ are described by the unperturbed (without taking
into account the Coulomb forces) wave functions, which are not
represented here.

Let us consider a pair, consisting of a magnetopolaron and hole.
In the case of the A magnetopolaron\cite{23} the following terms
are overcrossed: the exciton with the indexes  $n_{e}=1,n_{h}=1,m$
and exciton with indexes  $n_{e}=0,n_{h}=0,m$ plus the phonon with
the frequency $\omega_{LO}$. We have chosen $n_{h}=1,
m_{h}=m_{e}=m$, because such combination may be created by light
in the case of the infinitely deep QW. If one omits the correction
$\Delta_{1} {\cal E}_{n_{e},n_{h}}({\bf {\cal K}}_\perp)$, which
depends on  ${\bf {\cal K}}_\perp$, the resonant condition becomes
\begin{equation}
\hbar\omega_{eH}=\hbar\omega_{LO}+\Delta {\cal E}_{0,1}({\bf {\cal
K}}_\perp=0) -\Delta {\cal E}_{1,1}({\bf {\cal K}}_\perp=0).
\end{equation}
Because the excitonic corrections
  $\Delta {\cal E}_{0,1}({\bf {\cal K}}_\perp=0)$
and $\Delta {\cal E}_{1,1}({\bf {\cal K}}_\perp=0)$ are different
in values, the resonant condition Eq. (43) does not coincide with
the resonant condition Eq.
 (3), which has been obtained without taking into account
  the Coulomb forces. Applying \onlinecite{19},
one can obtain the equation for the energy ${\cal E}$ of the
magnetopolaron-hole pair
\begin{equation}
{\cal E}-{\cal E}_{1,1}({\bf {\cal
K}}_\perp)-\sum_{\nu}\frac{|U(\nu)|^2} {{\cal E}-{\cal
E}_{0,1}({\bf {\cal K}}_{\perp}-{\bf q}_{\perp})
-\hbar\omega_{LO}}=0.
\end{equation}
Inequalities of Eq. (38) and the estimates Eq. (39)  lead to the
hard restrictions of the QW width from above, what makes
problematic (in any case for GaAs) the application of the bulk
phonon approximation. Therefore  the index $\nu$ in Eq. (44)
includes the indexes ${\bf q}_\perp$ and $j$, where $j$ relates to
the confined and interface phonons. Because a dispersion of any
phonons is neglected in this section, the denominator in Eq. (44)
does not depend on  $j$.

Measuring the energy from the level
$$E_g+\varepsilon_m^e+\varepsilon_m^h+\frac{3}{2}\hbar\omega_{hH},$$
 one obtains
\begin{eqnarray}
{\cal E}_{1,1}({\bf {\cal K}}_\perp)=\Sigma_1+ \Delta {\cal
E}_{1,1}({\bf {\cal K}}_\perp),\nonumber\\ {\cal E}_{0,1}({\bf
{\cal K}}_\perp)=\Sigma_0+ \Delta {\cal E}_{0,1}({\bf {\cal
K}}_\perp).
\end{eqnarray}

Obviously, that the energy corrections $\Delta_{1} {\cal
E}_{n_{e},n_{h}}({\bf {\cal K}}_\perp)$ lead to the two essential
results. First, the energy of the magnetopolaron-hole pair begins
to depend on the value of the vector ${\bf {\cal K}}_\perp$. At
the normal light incidence
 $\Delta_{1} {\cal E}_{n_{e},n_{h}}({\bf {\cal K}}_\perp)=0$,
but the energy dependence on ${\bf {\cal K}}_\perp$ must appear at
the oblique light incidence on the QW surface. Second, the term
 ${\cal
E}_{0,1}({\bf {\cal K}}_{\perp}-{\bf q}_{\perp})$ in the
denominator ${\cal E}_p$ of the LHS Eq. (44) must lead to the same
qualitative results as the phonon dispersion, i. e. to the
additional shifts of the energies of the upper and lower polaron
levels and to the additional contributions into the inverse
lifetimes of the polaron states. To obtain more precise results
one have to take into account simultaneously the phonon dispersion
and Coulomb forces.

\section{Acknowledgements}
        S.T.P thanks the Zacatecas Autonomous University and the National
Council of Science and Technology (CONACyT) of Mexico for the
financial support and hospitality.
       This work has been partially supported by the Russian
Foundation for Basic Research and by the Program "Solid State
Nanostructures Physics".

\newpage
\begin{figure}
 \caption{The function $F(E)$, determining the position of the
 magnetopolaron energy levels (Eq. (19)) as a function of energy $E$
 under condition Eq. (3). $\beta_0=d/l=7.5,~~
 l=(\hbar/2m_e\omega_{LO})^{1/2}$, the size-quantization quantum
 number $m=1$; $c=0$ (curve 1), $c=0.04 gl^2$ (curve 2), $c=0.2 gl^2$
 (curve 3), $g=\alpha^{1/2}\hbar\omega_{LO}$. $E_a^\prime , E_b^\prime,
  E_c$ designate the position of the energy levels when the
  phonon dispersion exists; $E_a , E_b$
designate the position of the energy levels when the
  phonon dispersion is absent.}
\end{figure}
\begin{figure}
 \caption{The schematic view of the magnetopolaron energy levels
 $E_a^\prime , E_b^\prime,
  E_c$ and of the function $\Sigma_0+\hbar\omega_{LO}(q)$.
   Fig. 2a corresponds to the lesser
 phonon dispersion than Fig. 2b.}
\end{figure}
\begin{figure}
\caption{The position and broadening of the magnetopolaron energy
levels under the resonant conditions due to the phonon dispersion
as functions of the dispersion parameter $c$. $\beta_0=d/l=7.5,$ ,
the size-quantization quantum
 number $m=1$;
  $g=\alpha^{1/2}\hbar\omega_{LO}$. $E_a^\prime , E_b^\prime$
  designate the position of the energy levels when the
  phonon dispersion exists; $E_a , E_b$
designate the position of the energy levels when the
  phonon dispersion is absent. $\gamma_b$ is the inverse lifetime of
  the lower energy level due to the phonon dispersion.}
\end{figure}


\begin{references}
\bibitem[\ddag]{byline2} The main contributions - radiative and
non-radiative - have been calculated in\onlinecite{2}.
\end{references}
\end{document}